# Local structure and phonon states mediated by intercalation-driven doping in superconducting $Li_{1.0}(C_5H_5N)_yFe_{2-z}Se_2$

Alexandros Deltsidis,[a,b] Myrsini Kaitatzi,[a,b] Laura Simonelli,[c] Chris Stock,[d] David Voneshen,[e,f] and Alexandros Lappas[a,*]

[a]*Institute of Electronic Structure and Laser, Foundation for Research and Technology–Hellas, Vassilika Vouton, 71110 Heraklion, Greece*

[b]*Department of Materials Science and Engineering, University of Crete, Voutes, 70013 Heraklion, Greece*

[c]*ALBA Synchrotron Light Source, Carrer de la Llum 2-26, 08290 Cerdanyola del Vallés, Spain*

[d]*School of Physics and Astronomy, University of Edinburgh, Edinburgh EH9 3JZ, United Kingdom*

[e]*ISIS Facility, Rutherford Appleton Laboratory, Chilton, Didcot, Oxfordshire OX11 0QX, United Kingdom*

[f]*Department of Physics, Royal Holloway University of London, Egham TW20 0EX, United Kingdom*

[*] e-mail: lappas@iesl.forth.gr

# ABSTRACT

Intercalation of two-dimensional (2D) iron chalcogenides with molecular species requires disentangling electronic and structural contributions to understand the puzzling limit to superconducting transition temperature ($T_c$) at the frontier of long interlayer separations. Here, synchrotron X-ray absorption spectroscopy (XAS) at the Se K-edge sheds light on the impact of carrier-doping on the local structure of the high-$T_c$ (~39 K) $Li_{1.0}(C_5H_5N)_yFe_{2-z}Se_2$ phase. This material is derived by annealing the structurally related as-made derivative ($T_c$~ 44 K), with layers being primed apart by [alkali-molecule] guests. Metrics, such as, a reduced filling of Se 4$p$ orbitals and shorter Fe-Se bonds in the annealed phase, corroborate to a lower electron doping level with respect to the as-made one. Analysis of the metal-ligand thermal motion, based on the correlated Debye model, further relates the higher $T_c$ intercalates with the softening of the local Fe-Se bond. Beyond electronic effects, intercalation brings forth host-guest interactions that mediate the dynamics of the bulk crystal structure. For this, neutron time-of-flight spectroscopy on the annealed derivative, corroborates to the Se-Fe-Se layer being sensitive to chemical pressure effects imposed by the confined organic guests. This reflects in the phonon density of states, where harder low-energy transverse acoustic matrix phonons and molecular vibrations are witnessed, with respect to the pristine inorganic ($\beta$-FeSe) and organic ($C_5D_5N$) counterparts. On cooling through $T_c$, these excitations arrive without a collective magnetic-resonance mode – essential in unconventional, spin-mediated mechanisms – enquiring about deviations from optimal doping. The work highlights that when the Fe-square planes are tuned far apart, carrier-doping leveraged by intercalation plays a key role in the $T_c$ parametrization.

**Keywords:** layered superconductors, 2D materials, intercalation, electron doping, phonons, spin-resonance, XANES, EXAFS, inelastic neutron scattering

# I. INTRODUCTION

Iron-based superconductors (FeSCs) are a family of materials capable of achieving high critical transition temperatures ($T_c$), with single-layer FeSe films reaching a record of 65 K [1,2] compared to just 8 K in bulk β-FeSe [3]. High-$T_c$ superconductivity in these systems is driven by a range of unconventional behaviors due to strong electronic correlations [4]. The multiorbital and multiband nature of FeSCs influences the correlation strength mediated by doping (e.g., chemical substitution) and provides a setting for Mott physics [5]. Effectively, in their five-orbital model, when converted into a multiband electronic structure, electron and hole pockets appear at the Fermi surface, establishing the conditions for pairing in doped FeSCs. The pairing gap symmetry and structure arise from a complex interplay between intra-pocket (cf., Coulomb repulsion) and inter-pocket (cf., electron-hole correlations, enhanced by spin/charge fluctuations) interactions that are strongly modulated by carrier doping [6]. This leads to intricate phase diagrams and ordered phases [7] that experimentally can be engineered by adjusting chemical or structural properties through doping (or pressure), transforming a magnetic, non-superconducting parent compound into a superconducting, non-magnetic state. Research efforts, devoted in the parametrization of the $T$c, support a picture where maximal $T$c emerges at optimal inter-pocket interactions determined by electron (or hole) doping [8].

In view of the important role of doping, alternatives to facilitate such a capability are sought. Intercalation chemistry appears with a great potential in that respect because the weak interlayer van der Waals bonding between the FeSe layers makes the parent β-FeSe an ideal host. Alkali metals or solvates of alkalis with bases, such as ammonia, aliphatic and aromatic amines, can be inserted in-between the layers, fostering layered structures with $T$c as high as ~46 K [9]. In addition to separating the active layers and reducing their coupling, the co-intercalated alkali metal and molecular species (i.e., guests or spacer layer) act as charge reservoirs, capable to withdraw or inject carriers to the layered iron-chalcogenide (FeCh) host [10]. Depending on the guest species and the chemistry involved, the host can access a large gallery of interlayer separation, $d$ (i.e., the normal Fe-sheet distance between adjacent FeSe layers along the $c$-axis). The increased $T_c$ in such expanded-lattice phases has sparked the intriguing idea that widening the spacing between FeSe layers enhances the two-dimensionality of the band structure, thereby improving Fermi surface nesting – an element crucial to theories about the superconducting pairing mechanism [8,11].

Nevertheless, while $T_c$ is shown to scale linearly with $d$, a puzzling saturating response appears at $d > 8.6$ Å [12]. The apparent threshold-interlayer spacing seems to mark the phase-space where the Fermi surface becomes entirely 2D [13,14], hence pushing the layers beyond doesn't raise $T_c$ further (>46 K). However, the intercalation chemistry of such expanded-lattice Fe-chalcogenides ($d > 8.6$ Å), as for example, with either with alkali [15–17] or alkaline-earth [18] ammonia solutions, affords a variety of intermediate phases that arrive with a notable variation in the magnitude of the $T_c$. This enquires about the critical role of doping brought by the chemical constituents of the guests and its impact on the electronic correlations, as a parameter to leverage the $T_c$ magnitude when the layers are tuned to be very far apart.

Beyond ammonia, larger electron-donating molecules, such as pyridine ($C_5H_5N$ = $PyH_5$), can also be co-intercalated with Li in β-FeSe [19]. Challenges related to the growth of this complex system have recently been addressed by in situ synchrotron X-ray total scattering studies [20]. As with the $NH_3$-intercalated derivatives, and depending on the synthetic conditions, different polytypes may be obtained. In particular, annealing the as-made $Li_x(C_5H_5N)_yFe_{2-z}Se_2$ phase affords a variant with significantly shorter $d$ ($d_{as\text{-}made}$ ~ 16 Å vs. $d_{annealed}$ ~ 11 Å). The as-made and annealed derivatives of the $Li_x(C_5H_5N)_yFe_{2-z}Se_2$ belong also to the expanded lattice ($d > 8.6$ Å) Fe-based superconductors that carry variable $T_c$ (cf., $T_c^{as-made}$ ~ 44 K vs. $T_c^{annealed}$ ~ 39 K). As that, they are a good testing ground to assess differences in their (atomic and electronic) structure and dynamics to help delineate parameters that mediate the $T_c$ magnitude at the frontier of large interlayer separations.

As charge-transfer is commonly reflected on nearest-neighbor atom-atom bonds [21], element-selective X-ray absorption spectroscopy (XAS) offers a favorable probe to shed light on the evolution of local structure correlations with carrier doping. Hence, it has been utilized here to draw differences between the two $PyH_5$-intercalated polytypes. From the point of view of the Se K-edge, X-ray absorption near-edge structure (XANES) was able to assess the filling of the ligand available electronic states, while extended X-ray absorption fine structure (EXAFS) to acquire quantitative information on the local Fe-Se atom-atom correlations. In the presently studied expanded-lattices, XAS shows that the nearest-neighbor Fe-Se bond elongation and the associated softening of the correlated thermal motion, depicted in the Debye model, are identifying features of the raised degree of doping that brings forth noticeable differences in the $T_c$ of the two $PyH_5$-intercalated compounds.

Intercalation, however, outside electronic changes (cf., redox charge-transfer), as witnessed by the local structure, brings forth host-guest interactions that are expected to affect the bulk lattice dynamics. For this, we employed time-of-flight inelastic neutron scattering (INS), as a well-suited approach to assess not only the lattice but also the spin dynamics when the Se-Fe-Se sheets are primed far apart by [alkali-$PyH_5$] guests. Key aim to this end was to provide evidence on a collective neutron spin-resonance mode that has been claimed to arise from spin-fluctuations [22]. As its role in unconventional pairing mechanisms remains unresolved [11], it provides motivation to obtain experimental evidence on the momentum-resolved fluctuation spectrum. Such a resonant excitation has been observed before in the spectra of high-$T_c$ expanded-lattice, molecule-intercalated iron selenides [23,24]. However, on cooling below $T_c$ in the case of the annealed $Li_x(C_5D_5N)_yFe_{2-z}Se_2$, the resonance has remained unresolved, enquiring about deviations from optimal doping. Signatures of electron-lattice correlations, upon entering the superconducting state, are undistinguishable in the bulk material's phonon density of states with the current spectrometer resolution. However, certain low-energy acoustic vibrational modes become harder with respect to β-FeSe as they are sensitive to the raised compressibility of the bulk lattice due to the intercalated guests. The complementary views of structure and dynamics suggest that intercalation-induced charge-doping is essential for tweaking the electronic properties and parametrize $T_c$ when Fe-sheets are set far apart ($d > 8.6$ Å) in iron-based superconductors.

## II. METHODS

Synthesis of the parent β-FeSe and the as-made $Li_{1.0}(C_5H_5N)_yFe_{2-z}Se_2$ materials was reported previously [25]. For the synthesis of deuterated intercalated phases, an anhydrous reagent of 99.5% isotopic deuterium-substituted pyridine (Deutero GmbH) has been utilized, with additional drying over molecular sieves. As-made $Li_{1.0}(C_5H_5N)_yFe_{2-z}Se_2$ was annealed in a 5 mm quartz ampoule. The ampoule was pumped to $10^{-2}$ mbar, then charged with 600 mbar of He gas and flame-sealed under static conditions. The sealed ampoule was heated at 170 °C for 48 hours and then cooled to room temperature. This procedure was used to make both protonated ($C_5H_5N$) and deuterated ($C_5D_5N$) samples. All manipulations of materials were undertaken inside an Ar-circulating MBRAUN glovebox, with <1 ppm $O_2$ and $H_2O$. Basic characterization of the materials was performed by (i) powder X-Ray diffraction (PXRD), using a Bruker D8 Advance diffractometer with Cu-$K_a$ radiation, (ii) thermogravimetric analysis (TGA), with a TA Instruments SDT Q600 simultaneous TGA / DSC system, utilizing flowing Ar gas and (iii) magnetic measurements, with a Quantum Design MPMS XL7, superconducting quantum interference device (SQUID) magnetometer.

Inelastic neutron scattering (INS) measurements were performed on the time-of-flight chopper spectrometer MERLIN of the ISIS pulsed neutron facility [26]. Approximately 4 g of deuterated powder sample was loaded in an aluminum can, which was mounted on a top-loading closed-cycle refrigerator reaching down to 8 K. As the samples are highly air-sensitive, all handling was pursued in a high-quality He-circulating MBRAUN glove-box. Neutron inelastic scattering spectra were recorded as a function of momentum transfer, $Q$ and energy transfer, $E$ at various temperatures from 8 to 100 K. MERLIN's repetition-rate multiplication [RRM] mode has been utilized to allow simultaneous measurement of several incident energies. In the present work, neutron powder spectra $S(Q, E)$ were collected with $E_i$= 59.9 [24.5, 13.2] meV and $E_i$= 99.5 [22.8, 9.82] meV, with that of lower $E_i$ offering access to a narrower ($Q$, $E$)-space, but with higher energy-resolution. Data were reduced using Mantid [27] and the neutron-weighted phonon density of states (PDOS) obtained after correction for the Bose factor, $1/E$ and $Q^2$ terms.

Se ($E_0$ = 12658 eV) K-edge XAS measurements were performed in transmission mode at the beamline BL22-CLÆSS of the ALBA synchrotron [28]. The X-rays emitted by a multipole wiggler were monochromatized using a Si(311) double crystal monochromator, with Rh-coated mirrors used to reject higher harmonics. Samples loaded in an air-tight Al-container, with Kapton windows, were mounted on a continuous flow He cryostat in which the temperature was maintained within ±1 K through the 20-295 K range. Several scans were collected for any given temperature to ensure reproducibility and improve the signal-to-noise ratio. Data normalization and modelling (section S1) was performed with the DEMETER suite of programs [29].

## III. RESULTS AND DISCUSSION

### A. Basic characterization

Intercalation of [Li-$C_5H_5N$] moieties in the β-FeSe matrix ($d \sim 5.5$ Å) under modest temperature (80 °C), forms an expanded-lattice $Li_x(C_5H_5N)_yFe_{2-z}Se_2$ phase ($d \sim 16.2$ Å; Fig. 1a). Intense (00$l$) reflections of this so-called as-made phase, are indexed on the basis of $ThCr_2Si_2$ (122) structure type [20]. After annealing, the layered structure remains the same but the interlayer distance becomes shorter ($d \sim 11.4$ Å; Fig. 1a). Annealing the as-made $Li_{1.0}(C_5H_5N)_yFe_{2-z}Se_2$ compound reduces the $T_c$, from ~44 K to ~39 K (Fig. 1b). However, the superconducting transition becomes sharper (Fig. 1b,c). Moreover, annealing enhances the superconducting volume fraction with concomitant reduction of the normal state susceptibility (Fig. 1b). The deuterated annealed $Li_{1.0}(C_5D_5N)_yFe_{2-z}Se_2$ material carries the same $T_c$ as its protonated counterpart (Fig. S1a). From AC susceptibility measurements under different DC magnetic fields, up to 20 kOe, the onset of $T_c$ systematically shifts to lower temperatures as the applied DC field increases (Fig. S1a,b). Even when $H_{dc} = 10$ Oe, the $T_c$ shifts to ~37 K and the transition becomes broader (Fig. S1a), suggesting a low $H_{c1}$ (< 10 Oe), reminiscent of the $K_xFe_{2-y}Se_2$ and the $Rb_xFe_{2-y}Se_2$ materials, where $H_{c1} \cong 3$ Oe [30,31]. The shape of the $M$-$H$ loop measured at 5 K (Fig. S2) is typical of a type-II superconductor. The content of the molecule ($y$) in the $Li_{1.0}(C_5H_5N)_yFe_{2-z}Se_2$ formula is estimated by the weight loss curves of TGA (Fig. 1d) [25]. The pyridine content is calculated to be $y \sim 0.45$ and $y \sim 0.22$ in the as-made and annealed derivative, respectively. The experimentally calculated molecule contents ($y$) imply ~2 and ~1 pyridine molecules intercalated between the FeSe layers of the as-made and annealed derivatives, respectively [25]. This agrees with the formal size of a pyridine molecule (~5 Å), and is consistent with the interlayer separation of each derivative as conferred by PXRD (Fig. 1a and Fig. 2b,c).

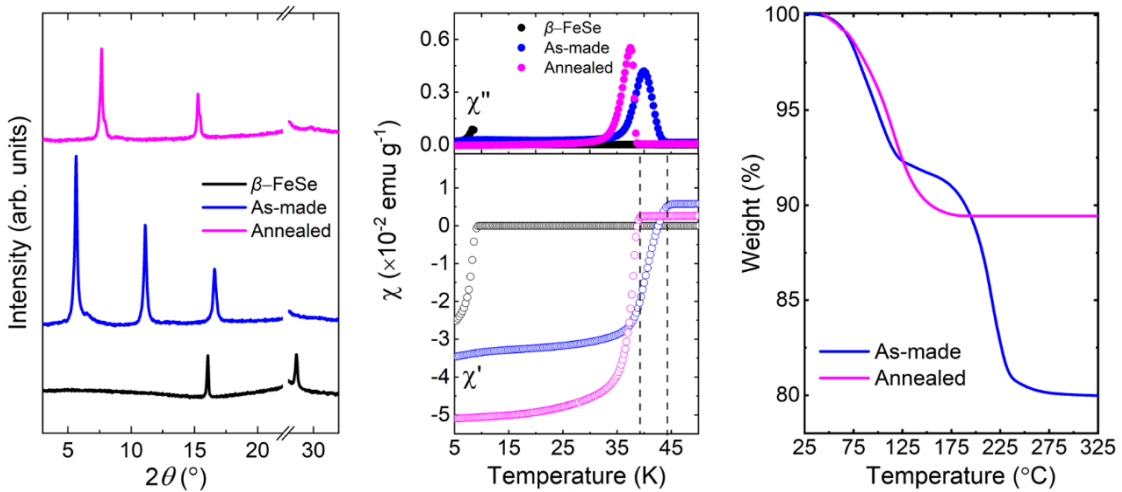

**Figure 1.** (a) PXRD (Cu K$\alpha$) patterns of the parent β-FeSe, as-made and annealed $Li_{1.0}(C_5H_5N)_yFe_{2-z}Se_2$ derivatives. The region of the PXRD corresponding to $2\theta = 22\text{-}28$ ° is excluded because of a contribution from the custom-made, air-tight sample holder. Normalized AC susceptibility ($H_{ac} = 1$ Oe and $f = 999$ Hz) of the (b) real ($\chi'$) and (c) imaginary ($\chi''$) parts of the parent β-FeSe, as-made and annealed $Li_{1.0}(C_5H_5N)_yFe_{2-z}Se_2$ derivatives. (d) TGA weight-

loss curves, under Ar gas flow (100 ml/min), for the as-made and annealed $Li_{1.0}(C_5H_5N)_yFe_{2-z}Se_2$ compounds.

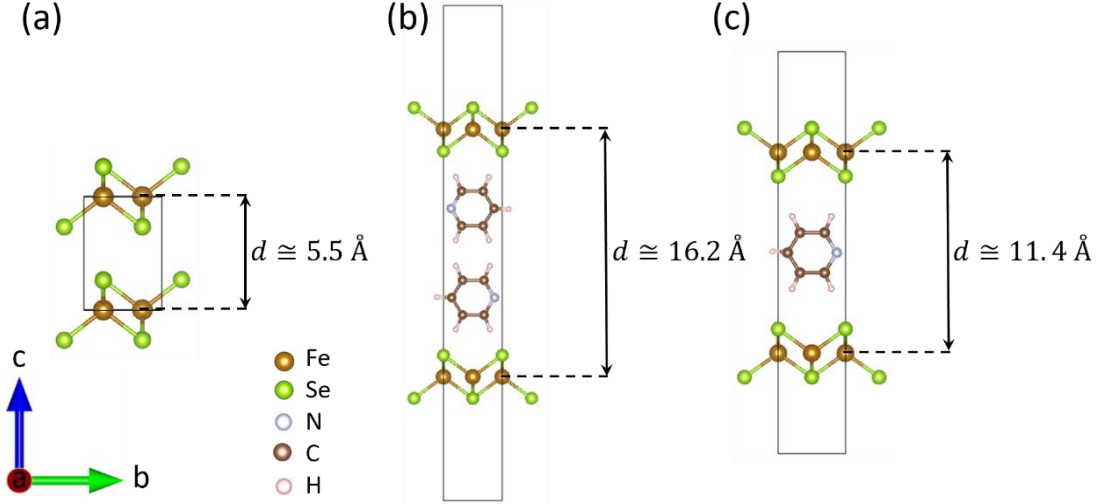

**Figure 2.** A schematic view depicting the interlayer separation ($d$) across the FeCh phases: parent β-FeSe (a); $Li_{1.0}(C_5H_5N)_yFe_{2-z}Se_2$, entailing the as-made (b) and annealed (c) derivatives.

## B. Inelastic neutron scattering

INS data of the annealed $Li_{1.0}(C_5D_5N)_yFe_{2-z}Se_2$ compound were collected at several temperatures across the $T_c$. Representative neutron powder spectra, $S(Q,E)$ (Fig. 3a,b), measured with $E_i = 24.5$ meV, offer improved resolution in the low $E$ region and those with $E_i = 99.5$ meV, allow for an extended energy coverage. Besides scattering from phonons, careful inspection of the $S(Q,E)$ intensity maps could not identify a resonant magnetic excitation [11], localized in both energy ($E_{res}$) and wavevector ($Q_{res}$). However, such a signal has been found before to develop on cooling through $T_c$ in both the parent β-FeSe [32,33] as well as its alkali- [34,35] and molecule- [23,24] intercalated derivatives. In the various Fe-based polytypes, it arises from quasi-two-dimensional spin fluctuations that resonate at somewhat different in-plane wavectors, $Q_{res}$ (e.g., ~1.14 -1.24 Å$^{-1}$), with respect to the periodicity of the Fe-square sublattice 2D antiferromagnetic ordering [36]. Since this spin-resonance scales with the $T_c$ (cf., $E_{res} \approx 5.8k_B T_c$), it has been considered as a key for testing the unconventional symmetry of the superconducting order parameter [37].

In $Li_{1.0}(C_5D_5N)_yFe_{2-z}Se_2$, though, the strong phonon signals may hinder a possible column of scattering characteristic of fluctuating magnetic moments, in the range of relevant wavevectors [e.g., (π,0)-type at $Q_{res} \sim 1.2$ Å$^{-1}$]. In addition, while resonant spin-excitations in the superconducting state likely stem from electron scattering between Fermi surface regions with opposite order parameter signs [11], deviation from optimal doping (where maximal $T_c$ is attained), can alter electron/hole pockets and weaken inter-band transitions, reducing resonant spin-fluctuations [6]. In that respect, the possibility of the system residing away from the optimally-doped regime [37], may not be ruled out as a reason for the lack of a spin-resonance mode in this molecule intercalated phase. Nevertheless, the PDOS, in the superconducting ($T = 8$ K) and at the normal states ($T$ = 49 K), reveals numerous features labeled as (1) – (11) in Fig. 3c-e. Previous INS experiments on the parent β-FeSe, showed that there are eight phonon modes below the cutoff energy of 40 meV [38,39]. Here, due to limited instrumental resolution only seven modes, (1) – (7) (Fig. 3c,d), can be resolved and assigned to vibrations originating from the FeSe$_4$ building blocks of the layered FeSe host. Moreover, in the energy range 40 – 80 meV, the modes (8) – (11) (Fig. 3d,e) assign well to vibrations characterizing the intercalated molecular guest, C$_5$D$_5$N [40,41].

Figure 3c shows the low-energy features assigned as: #1 and #2, are transversal acoustic (TA) modes [39,42] which are approximately of equal mixture of Fe and Se contributions [42]; #3 and #4, are longitudinal acoustic (LA) modes [39], both dominated by Se vibrations [42]. Their temperature evolution is compiled in Fig. S3a. Figure 3d shows the medium-energy features, namely: #5 – #7 are optic modes that belong to the inorganic host [39] and are mainly of Fe character [42], while #8 and #9 are due to the intercalated molecule, assigned to the $\nu_{16a}$ and $\nu_{16b}$ type of ring vibrations of C$_5$D$_5$N [40]. Figure 3e presents the high-energy features, specifically: #10 is assigned to the $\nu_{11}$, and #11 to the $\nu_{6a}$ - $\nu_{6b}$ modes of the C$_5$D$_5$N guest [40]. Although #8 – #10 are out-of-plane ring vibrations, #11 corresponds to in-plane molecular motion [41,43]. All the modes of the intercalated C$_5$D$_5$N seen in the PDOS here (*cf*. #8 – #11, Fig. 3d,e), reside at somewhat higher energy with respect to the modes of the free C$_5$D$_5$N (Table 1) [40], a likely outcome of its confinement in the 2D layers of the host. With temperature lowering, the increased intensity of such features (Fig. S3b) suggests a somewhat reduced molecular disorder for their restricted dynamics. Moreover, the intensity-mode evolution for the guests, shows no abrupt changes at $T_c$ that would have otherwise provided evidence of coupling to the emerging electron-correlations upon entering the superconducting state.

Further examining the phonon modes of the electronically active FeSe layers provides useful insights. Between the superconducting ($T$ = 8 K) and the normal ($T$ = 49 K) states, there is subtle difference in the intensity of the modes belonging to the host. The low-energy TA modes #1 – #2 become slightly sharper on cooling from 100 K to base temperature (Fig. S3a), as it has been observed in the parent β-FeSe [39]. Within the current energy resolution, the modes of the inorganic framework are observed at energies comparable to the parent β-FeSe (Table 1) [38,39,42]. Exception, is the lowest energy TA mode (cf., #1, Fig. 3c) that is located around 6.7 meV, which is ~20% higher than that in the parent β-FeSe (Table 1) [38,39]. The evolution of this TA mode can be rationalized in view of Lifshitz's predictions for the so-called "bending-waves" in highly anisotropic layered crystals [44]. Such low-energy, out-of-plane modes play a key role in the phonon dispersions because of the small value of the interlayer elastic constants [45], and as such they are sensitive to the compressibility of the lattice.

In the case of parent β-FeSe, the soft van der Waals interlayer bonding facilitates enhanced *c*-axis compressibility. Given the softness of the interlayer Se-Se interactions [39,46], applied hydrostatic pressure raises $T_c$ (up to 37 K) and hardens the TA phonons as structural compaction strengthens the overall lattice cohesion [39]. Since low-energy acoustic phonons reflect the materials' bulk mechanical properties, like elasticity and stiffness [47], intercalation of [Li-C$_5$D$_5$N] in β-FeSe may assume to offer a related to structural compaction from external pressure channel for property tuning. Priming the host layers apart by inserting guest molecules in the van der Waals space, beyond electronic changes (cf., redox charge-transfer or modification of the ligand field), brings forth host-guest steric interactions that may lessen *c*-axis compressibility. The effect on the layered structure may be chemical pressure exerted by the guests on the interlayer Se-atoms. Then, structural compaction of the Se-Fe-Se sheets by analogy to the pressurized phase of β-FeSe, shifts sensitive out-of-plane (TA) vibrations of the intercalated phase to higher energy (cf., #1, Table 1).

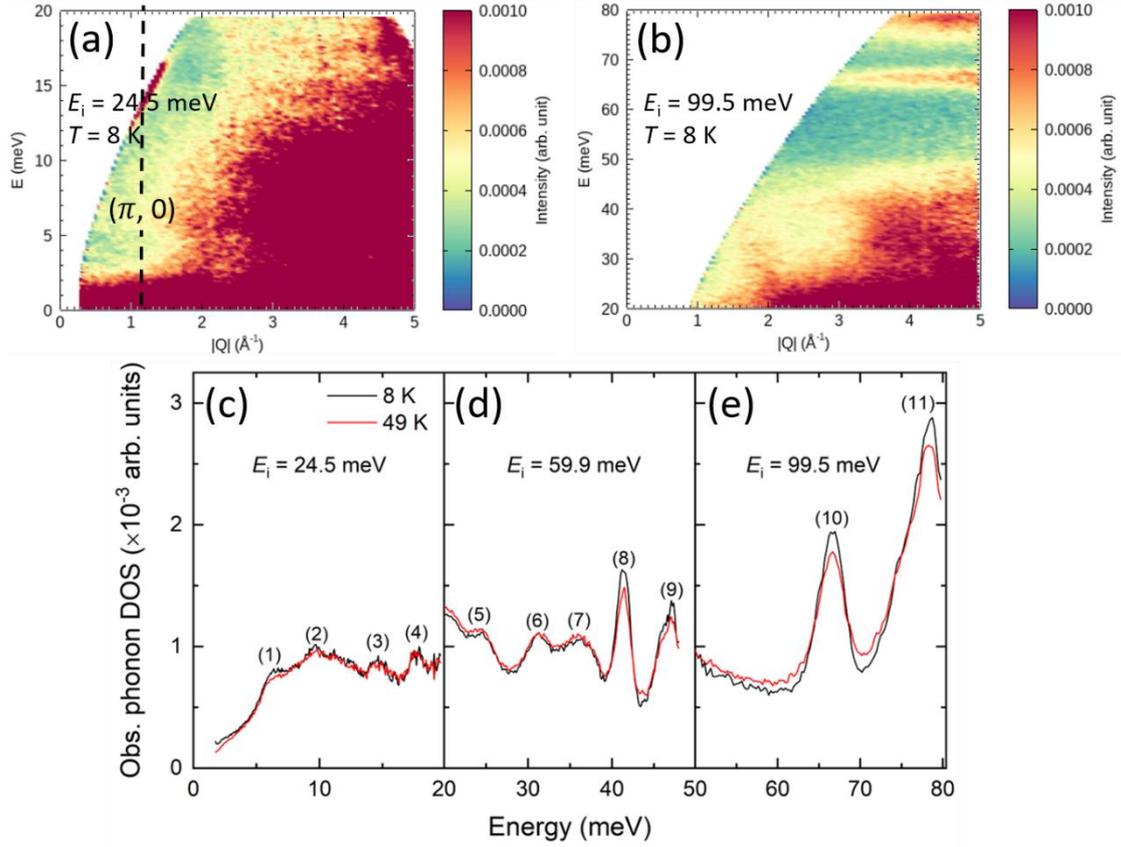

**Figure 3.** Neutron powder spectra, $S(\mathbf{Q}, E)$, of the annealed $Li_{1.0}(C_5D_5N)_yFe_{2-z}Se_2$ derivative at $T = 8$ K: (a) low-energy ($E_i = 24.5$ meV) and (b) high-energy ($E_i = 99.5$ meV) part of the spectrum. The vertical dashed line denotes the equivalent 2D wave vector of (π, 0)-type $(Q = 1.2 \text{ Å}^{-1})$ of scattering (see text). (c – e) The measured PDOS of the annealed $Li_{1.0}(C_5D_5N)_yFe_{2-z}Se_2$ at the superconducting ($T = 8$ K) and normal ($T = 49$ K) states. The PDOS are extracted from spectra with different incident energies, probing the: (c) low-energy ($E_i = 24.5$ meV), (d) medium-energy ($E_i = 59.9$ meV) and (e) high-energy ($E_i = 99.5$ meV) regions.

**Table 1**. Base-temperature phonon modes of the annealed $Li_{1.0}(C_5D_5N)_yFe_{2-z}Se_2$ compared with the pristine inorganic ($\beta$-FeSe) and organic ($C_5D_5N$) counterparts*. Mode labels: see Fig. 3c-e. TA, is transverse acoustic and LA, is longitudinal acoustic.

|  | Se-Fe-Se layer | | | | $C_5D_5N$ molecule | | | |
|---|---|---|---|---|---|---|---|---|
| Phonon modes | #1 | #2 | #3 | #4 | #8 | #9 | #10 | #11 |
| Pristine (meV) | 5.6 | 9.4 | 15-18 | | 39.4 | 45.6 | 65.2 | 77.6 |
| Intercalated (meV) | 6.7 | 9.7 | 14.5 | 17.7 | 41.3 | 47.7 | 66.6 | 78.6 |
| Assignment | TA | TA | LA | LA | $\nu_{16a}$ | $\nu_{16b}$ | $\nu_{11}$ | $\nu_{6b}$ |

* Vibrational energies and assignment: Se-Fe-Se layer, [39]; $C_5D_5N$ in gas phase, [40].

## C. X-ray absorption spectroscopy

### 1. XANES

XAS spectra of the annealed $Li_{1.0}(C_5H_5N)_yFe_{2-z}Se_2$ compound were collected at the Se K-edge, on cooling and then upon warming, between room temperature and 20 K. The XAS spectra do not show any major changes with temperature across $T_c$ (Fig. 4a). Figure 4b presents the normalized XANES of the annealed material—in the superconducting state at 20 K—compared with the respective spectra of the as-made and the parent β-FeSe compounds. The normalized Se K-edge spectra reveal the typical features of $Se^{2-}$ systems [48,49]. Above the absorption edge, at ~ 1 eV the spectra show a strong peak (feature #A; Fig. 4b), and at ~ 7 eV, a much broader feature (feature #B; Fig. 4b). The latter is governed by multiple scattering between the photoelectron and its neighbors therefore it is related to the local atomic environment. The former (#A) is due to the Se $1s \rightarrow 4p$ transition and it is related to the unoccupied density of states (DOS) near the Fermi level ($E_F$).

Feature #A modulates across the FeCh derivatives. From parent β-FeSe to the as-made derivative, decreases in intensity suggesting a reduced number of unoccupied Se $4p$ states near $E_F$; this implies that the as-made compound is electron doped compared to the parent. From the as-made to the annealed derivative, feature #A increases in intensity, suggesting that the number of unoccupied Se $4p$ states becomes larger when the as-made compound is annealed. The effect on the intensity of feature #A is clearly visible in the difference spectra (Fig. 4c; i.e., the parent β-FeSe spectrum is subtracted from the spectra of the as-made and annealed derivatives). The small difference in the intensity of #A, between the as-made and the annealed derivatives (Fig. 4c), implies a discernible difference in the DOS between the two polytypes. In the as-made derivative more electrons occupy the Se $4p$ states, suggesting band-filling with respect to that of the annealed derivative (Fig. 4d).

Feature #B also modulates across the FeCh derivatives and shows the opposite trend to feature #A. Compared to the parent β-FeSe, #B in the expanded lattices increases in intensity (Fig. 4b), indicating reorganization of the lattice. In an earlier study of the as-made $Li_x(C_5H_5N)_yFe_{2-z}Se_2$ as a function of Li content ($x$), it was shown that the intensity of feature #B increases systematically with increasing doping ($x$), while the corresponding Fe-Se bond length becomes longer [25]. Here, feature #B has lower intensity in the annealed phase as nicely reflected in the difference spectra (Fig. 4c). Hence, we argue that the reduced intensity of #B in the annealed (Fig. 4b) is correlated with shorter Fe-Se bond length. This aligns with the argument on a lower electron doping in the annealed derivative (*cf.*, feature #A, Fig. 4b,c; *vide supra*).

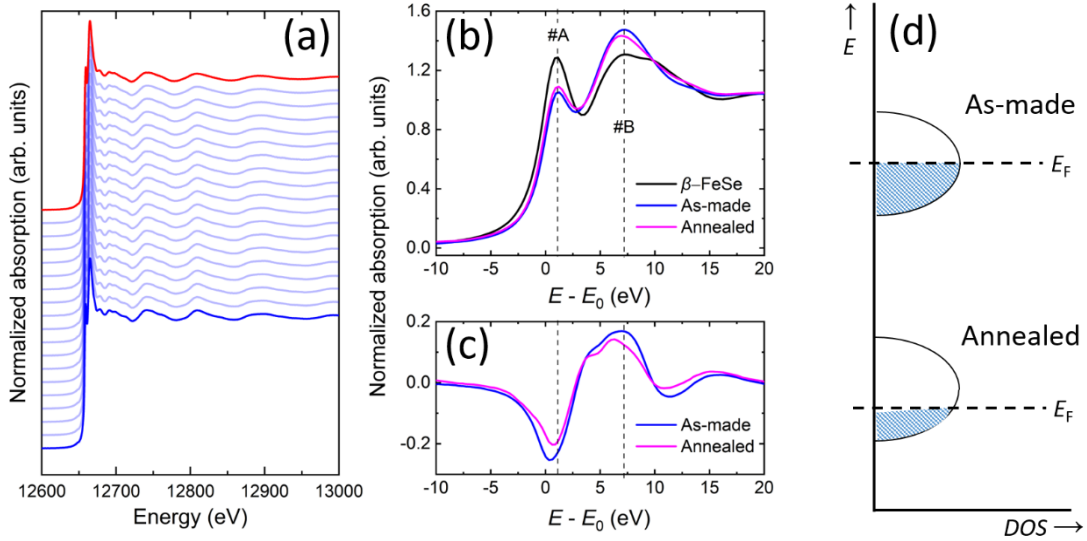

**Figure 4.** (a) Normalized XAS versus energy measured at the Se K-edge for the annealed $Li_{1.0}(C_5H_5N)_yFe_{2-z}Se_2$ as a function of temperature. The data are shifted vertically for clarity. Blue is data at 20 K, red is data at 295 K, and the light-blue colored lines are the intermediate temperatures. (b) Normalized XANES region of the Se K-edge XAS for the parent β-FeSe, as-made and annealed $Li_{1.0}(C_5H_5N)_yFe_{2-z}Se_2$ derivatives measured at 20 K and the corresponding (c) difference spectra (see the text). The data of β-FeSe and as-made $Li_{1.0}(C_5H_5N)_yFe_{2-z}Se_2$ are compiled from reference [25]. The *x* axis is presented after subtracting the normalized $E_0$ value of each spectrum from the respective incoming energy. (d) Schematic representation of the density of states of the Se 4*p* band for the as-made and annealed derivatives.

## 2. EXAFS

The XANES region at the Se K-edge confers discernible differences in the environment around the Se absorber of each FeCh derivative. These differences arise from electron doping effects that reflect in the nearest Fe-Se bond (cf., features #A and #B, Fig. 4b,c; *vide supra*). Such atom-atom correlations are better evaluated through the EXAFS region. Figure 5a shows the Se K-edge $k^2$-weighted EXAFS oscillations of the annealed $Li_{1.0}(C_5H_5N)_yFe_{2-z}Se_2$ compound. The evolution of the oscillations between $20 \leq T \leq 295$ K, with increasing *T* (Fig. 5a) is due to thermal damping. There are no signs of possible subtle structural phase changes when the EXAFS oscillations of the annealed are compared against the as-made derivative. However, close inspection of the EXAFS for the two polytypes, at a representative temperature (20 K), shows that those of the as-made are slightly compressed to lower *k* with respect to the annealed case (Fig. 5b). This implies that the Fe-Se bond length of the as-made should be longer than that of the annealed. The effect is consistent with the XANES features (#A, #B; Fig. 4b,c) that advise for a higher electron count in the as-made derivative.

For quantitative analysis, the Fourier transforms (FTs) of the Se K-edge $k^2$-weighted EXAFS oscillations were taken from the *k* range 3.55-15.1 Å$^{-1}$ with a Hanning window. Representative FTs at 20 K and 295 K are shown in Figure 5c. They are dominated by contributions from the 1$^{st}$ coordination shell of Se-Fe (inset, Fig. 5c) at ~2 Å. Peaks at longer R (>3 Å) are associated to mixed single and multiple scattering contributions from neighboring coordination shells. Contributions in the FTs (Fig. 5c) appear at shorter R than that expected from the crystallographic model because the FTs were not corrected for the phase shifts implemented in the backscattered wave. Quantitative information on the Se-Fe atom-atom correlations is extracted by fitting a structural model to the EXAFS oscillations (section S1). The fitting within the 1$^{st}$ coordination shell (inset, Fig. 5c) around the Se absorber is performed in the context of single-scattering approximation (section S1) by employing the tetragonal crystal structure (*P4/nmm*) of the layered β-FeSe [25].

## *2.1. Doping-driven local structure distortions.*

It is well established that redox processes, involving charge-transfer, can significantly affect metal-ligand bond lengths by altering the electronic configuration and bonding interactions between the metal center and the ligand [50]. For instance, reduction, brought about by electron-doping introduces electrons into antibonding states, causing increased electron-electron repulsion that can weaken the overall bonding in a material, and result in a longer bond length as atoms are less tightly bound [51]. Valuable in that sense is the analysis of the Se K-edge EXAFS in the present study, as it enables assessment of the local Fe-Se bond evolution across the FeCh derivatives. With respect to the parent β-FeSe, the Fe-Se bond of the as-made is elongated more than that in the annealed (cf., $\Delta R_{\text{Fe-Se}} \cong 0.017(2)$ Å *vs* $\Delta R_{\text{Fe-Se}} \cong 0.012(1)$ Å, respectively; Fig. 5d). This is consistent with the compressed (to lower *k*) EXAFS oscillations of the as-made (Fig. 5b) and the evolution of the XANES features (cf., #A and #B, Fig. 4b,c; *vide supra*).

Assessments against pioneering molecular intercalates of NH$_3$-related systems offer useful insights. The elongation of the local Fe-Se bond on going from [Li-PyH$_5$] -poor to -rich compositions (Fig. 5d; Table 2) follows the same trend as in the [Li-NH$_3$] intercalates [15,16]. However, with concomitant enhancements of *d* and the amount of the intercalated molecule (PyH$_5$ vs. NH$_3$), the opposite evolution to the magnitude of $T_c$ is witnessed (Table 2). In view of this, electronic structure calculations on the [Li-NH$_3$] intercalates interpret the $T_c$ depression with *d* to a reduced level of electron-doping due to a raised content of amide (NH$_2$) co-intercalated species that impede effective charge-transfer [13].

The amount of spacer molecules intercalated in FeCh compounds, though pushes the interacting layers apart and makes the Fermi surface acquiring a favorable 2D character [13,14], is not the only factor facilitating high $T_c$. Beyond that, the level of efficient carrier-doping brought by the chemical constituents of the guests can potentially advance the superconducting properties. This is demonstrated here in the [Li-PyH$_5$] phase-space, where on going from the annealed to the as-made phases, the Fe-Se bond lengthening (Table 2) confers a raised level of electron doping [51] that aligns with maximal $T_c$ (Fig. 1b).

**Table 2.** Interlayer spacing (*d*), critical temperature ($T_c$) and Fe-Se bond length ($R_{\text{Fe-Se}}$) for the [Li-NH$_3$] and [Li-C$_5$H$_5$N] intercalated FeSe superconducting phases. The $R_{\text{Fe-Se}}$ bond length of the NH$_3$-poor and rich derivatives are at 5 K. The $R_{\text{Fe-Se}}$ bond length of the C$_5$H$_5$N derivatives are extracted from the Li$_x$(C$_5$H$_5$N)$_y$Fe$_{2-z}$Se$_2$ EXAFS analysis at 20 K (see text).

|  | NH$_3$-poor [15] | NH$_3$-rich [16] | C$_5$H$_5$N-poor (annealed) | C$_5$H$_5$N-rich (as-made) |
|---|---|---|---|---|
| *d* (Å) | 8.5 | 10.6 | 11.4 | 16.2 |
| $T_c$ (K) | 43 | 39 | 39 | 44 |
| $R_{\text{Fe-Se}}$ (Å) | 2.408(1) | 2.415(1) | 2.3908(8) | 2.394(1) |

## *2.2. Correlated local structure disorder.*

Identifying changes in the local bonding of the Fe-Se atoms is expected to affect the respective atom-atom correlations. One effective way to describe this is through the mean-square relative displacements (MSRDs; Debye-Waller factors). They include contributions from the temperature-independent ($\sigma_s^2$; static) and -dependent ($\sigma_d^2$; dynamic) terms: $\sigma^2 = \sigma_s^2 + \sigma_d^2(T)$ [52]. Analysis of the Se K-edge EXAFS uncovers the Fe-Se MSRDs. Those of the annealed are systematically lower than those of the as-made (Fig. 5e), with no anomaly across $T_c$ within the error bars. Furthermore, data collected on warming (Fig. 5e) —for the annealed derivative—imply that the nature of the sample is not glassy [53,54]. Interestingly, the MSRDs of the annealed are similar to those of the parent β-FeSe in the full temperature range (inset, Fig. 5e). This would suggest similar Fe-Se atom correlations for the two systems. However, the parent β-FeSe and the annealed compounds have very different $T_c$s, and more importantly a dissimilar Fe-Se bond length ($R_{\text{Fe-Se}}$, Fig. 5d), which reflects varying electron doping effects [51,55]. Phenomenological models can be fitted against the MSRDs to assess correlated effects in the fluctuations of interatomic distances due to thermal evolution. The MSRDs extracted from the EXAFS analysis can be modelled either on the basis of the correlated Einstein or Debye models (section S1) [56].

Analysis with the Einstein model finds similar, within the error bars, Einstein temperature ($\theta_E$, Table 3), suggesting a comparable local lattice force constant, $k \sim 5.9 \pm 0.1$ eV Å$^2$ ($k = \mu\omega_E^2$, Section S1) amongst the derivatives. On the other hand, the temperature-independent term ($\sigma_s^2$) differs significantly between the as-made and annealed derivatives. The almost four times lower static disorder (Table 3) in the annealed could imply local reorganization of Fe atomic sites. In earlier work, on a series of as-made Li$_x$(C$_5$H$_5$N)$_y$Fe$_{2-z}$Se$_2$ derivatives with varying doping, it was shown that in the $x = 1.0$ compound, the FeSe layer involved fewer than 4 Fe atoms within the 1$^{\text{st}}$ coordination shell around the Se absorber (inset, Fig. 5c) [25]. Here, refining the coordination number (*N*) of nearest-neighbor Fe atoms around the Se absorber suggests that the annealed is stoichiometric (Table 3), in line with its lower static disorder. Therefore, annealing seems to heal the Fe-site vacancies in the FeSe slabs, in accord with the sharp transition at $T_c$ and a homogeneous superconducting phase (Fig. 1b,c).

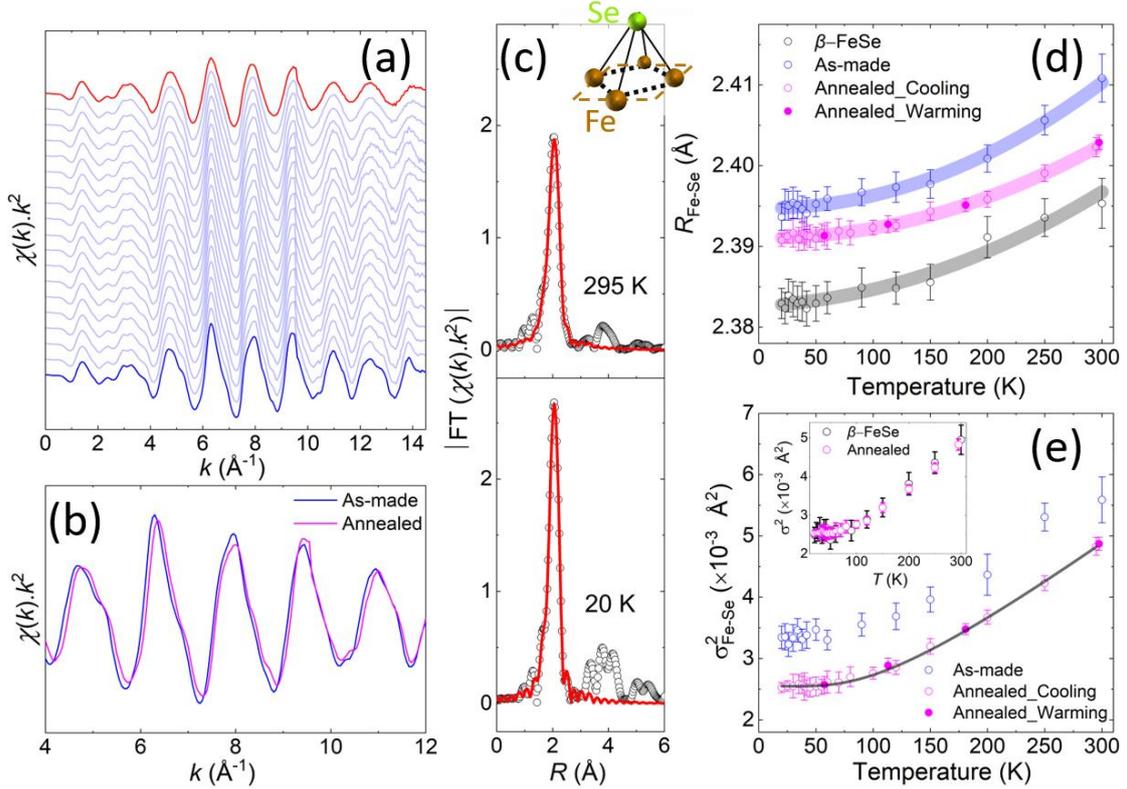

**Figure 5.** (a) Se K-edge EXAFS oscillations of the annealed $Li_{1.0}(C_5H_5N)_yFe_{2-z}Se_2$ as a function of temperature. The data are shifted vertically for clarity. Blue-line, is data at 20 K, red-colored at 295 K, and the light-blue lines are the intermediate temperatures. (b) Se K-edge EXAFS oscillations of the as-made (blue) and annealed (magenta) derivatives at 20 K. (c) Representative fits of the FTs of the annealed $Li_{1.0}(C_5H_5N)_yFe_{2-z}Se_2$; the data and model are shown with open circles and solid lines, respectively. Inset: Schematic of the 1$^{st}$ coordination shell around the Se absorber; green ball, Se atom and brown ball, Fe atom. Temperature evolution of the local structure parameters in the FeCh compounds: (d) Fe-Se pair bond distance and (e) Fe-Se pair MSRDs. The data of β-FeSe and as-made $Li_{1.0}(C_5H_5N)_yFe_{2-z}Se_2$ are taken from our previous study [25]. The bold lines in (d) are guides to the eye. The solid line in (e) is the least-squares fit based on the correlated Einstein model (see text).

It is worth noting though that the bulk lattice properties, as reflected in the neutron PDOS upon intercalation of [Li-$C_5D_5N$] moieties in the β-FeSe matrix, display multiple phonon modes (Fig. 3 and S3). In view of this, the Einstein model that describes the whole spectrum with a single vibrational frequency, seems to offer only a basic approximation of the relevant energy scale. To attain a somewhat improved assessment of the local metal-ligand bond strength, we employed the correlated Debye model (section S1) that approximates the spectrum linearly through a distribution of frequencies [56]. The subsequent analysis of the MSRDs now suggests differing Debye temperatures ($\theta_D$) amongst the FeCh derivatives (Table 3). In effect, as the interlayer separation, $d$, increases, $\theta_D$ is lowered, suggesting a reduced stiffness of the bonding between metal-chalcogen at the local level. A similar trend is also observed upon molecule-based intercalations of superconducting layered $TaS_2$ [57]. Such a softening of the local bond often results from electronic changes, such as charge transfer or modifications in the ligand field [51]. In that respect, the local metal-chalcogen correlations picked up by the Debye model suggest that the Fe-Se bond becomes softer ($\theta_D$, Table 1) with increasing interlayer spacing and concomitant charge-doping, as it is corroborated by the EXAFS (cf., elongation of the Fe-Se bond itself; Fig. 5d) and the XANES (cf.; filling of Se 4$p$ orbitals; Fig. 4b-d).

**Table 3.** Interlayer spacing ($d$), critical temperature ($T_c$) and parameters extracted from the Se K-edge EXAFS analysis of the Li$_x$(C$_5$H$_5$N)$_y$Fe$_{2-z}$Se$_2$ data (see the text). Analysis of the mean-square relative displacements on the basis of the correlated Einstein and Debye models provided the Einstein ($\theta_E$) and the Debye ($\theta_D$) temperatures, as well as the static disorder ($\sigma_s^2$) (see text). The coordination number ($N$) was obtained by fitting a structural model to the 20 K data.

| FeCh system | $d$ (Å) | $T_c$ (K) | $\theta_E$ (K) | $\sigma_s^2 \times 10^{-4}$ (Å$^2$) | $N$ | $\theta_D$ (K) |
|---|---|---|---|---|---|---|
| Parent | 5.5 | 8.5 | 318.2 ± 4.8 | 2.1 ± 0.5 | 3.97 ± 0.11 | 380.0 ± 5.3 |
| Annealed | 11.4 | 39 | 325.0 ± 1.3 | 2.7 ± 0.1 | 4.01 ± 0.07 | 368.4 ± 3.2 |
| As-made | 16.2 | 44 | 319.4 ± 5.0 | 10.3 ± 0.5 | 3.88 ± 0.09 | 312.1 ± 4.4 |

## IV. CONCLUSIONS

In summary, the temperature dependent Se K-edge XAS is utilized to investigate the electronic and local structures of the high-$T_c$ Li$_{1.0}$(C$_5$H$_5$N)$_y$Fe$_{2-z}$Se$_2$ superconductor, derived after annealing its structurally related as-made compound. The $T_c$ lowering in the annealed derivative (from ~44 K to ~39 K) suggests a reduced electron doping level. This is shown through the smaller filling of the corresponding Se-ligand orbital states and the shorter Fe-Se nearest-neighbor bond, verified by XANES and EXAFS, respectively. The accompanying Fe-Se thermally assisted atomic fluctuations, approximated by the correlated Debye model, point that the raised $T_c$ in the intercalated derivatives, with increasing interlayer spacing, comes along with a softer Fe-Se local bonding. Except the afore-mentioned consequences due to charge-transfer effects, intercalation brings forth host-guest interactions that impact the bulk lattice dynamics. Effectively, the Se-Fe-Se layer becomes sensitive to chemical pressure effects imposed by the intercalated [Li-C$_5$D$_5$N] guests. This is reflected in the phonon density of states of the annealed Li$_{1.0}$(C$_5$D$_5$N)$_y$Fe$_{2-z}$Se$_2$ as probed by neutron time-of-flight spectroscopy. Characteristic is the hardening of low-energy, transverse acoustic phonons, which confer a reduced compressibility with respect to the lower $T_c$, parent β-FeSe. On cooling below $T_c$, resonant spin-fluctuations, claimed to act as the pairing glue in unconventional superconductivity, have remained unresolved, enquiring about deviations from optimal doping in the annealed Li$_x$(C$_5$D$_5$N)$_y$Fe$_{2-z}$Se$_2$. The outcomes highlight that in the expanded-lattice molecule-intercalated Fe-superconductors, when layers are tuned far apart ($d > 8.6$ Å), the guests' chemical constituents' sensitively mediate carrier-doping and electronic correlations to leverage the $T_c$ magnitude.

**ACKNOWLEDGEMENTS**

This material is based upon research supported by the Office of Naval Research Global under award no. N62909-17-1-2126. The synchrotron X-ray spectroscopy experiments were performed at BL22-CLÆSS beamline at ALBA Synchrotron (Barcelona, Spain) with the collaboration of the ALBA staff. Beamtime at BL22 was awarded after evaluation of proposal with ID 2023027395. Experiments at the ISIS Pulsed Neutron and Muon Source were supported by a beamtime award from the Science and Technology Facilities Council [58]. Deuterated pyridine ($C_5D_5N$) was provided by the ISIS deuteration facility.

**Supporting Information:** The Supporting Information is available free of charge at

Additional methods for experiments; details on XAS normalization, XAS modelling, AC susceptibility as a function of applied DC magnetic field, hysteresis loop at 5 K and PDOS as a function of temperature (DOC).

# Local structure and phonon states mediated by intercalation-driven doping in superconducting Li$_{1.0}$(C$_5$H$_5$N)$_y$Fe$_{2-z}$Se$_2$

Alexandros Deltsidis,[a,b] Myrsini Kaitatzi,[a,b] Laura Simonelli,[c] Chris Stock,[d] David Voneshen,[e,f] and Alexandros Lappas [a,*]

[a]*Institute of Electronic Structure and Laser, Foundation for Research and Technology–Hellas, Vassilika Vouton, 71110 Heraklion, Greece*

[b]*Department of Materials Science and Engineering, University of Crete, Voutes, 70013 Heraklion, Greece*

[c]*ALBA Synchrotron Light Source, Carrer de la Llum 2-26, 08290 Cerdanyola del Vallés, Spain*

[d]*School of Physics and Astronomy, University of Edinburgh, Edinburgh EH9 3JZ, United Kingdom*

[e]*ISIS Facility, Rutherford Appleton Laboratory, Chilton, Didcot, Oxfordshire OX11 0QX, United Kingdom*

[f]*Department of Physics, Royal Holloway University of London, Egham TW20 0EX, United Kingdom*

[*] e-mail: lappas@iesl.forth.gr

# S1. Methods

**X-Ray Absorption Spectroscopy: normalization and modelling**

The raw Se K-edge XAS spectra have been normalized within the Athena software [S1]. Initially, the threshold energy $E_0$ was approximated as the energy of the maximum derivative of the absorption spectra, $\mu(E)$. Then, a smooth pre-edge line was fitted to the pre-edge region to remove instrumental background, and a quadratic polynomial fit was fitted to the post-edge region to normalize the spectra to the absorption of an isolated atom.

Modelling of the EXAFS oscillation was performed within the Artemis software [S1]. The $k^2$-weighted EXAFS oscillations have been modelled on the basis of the single scattering approximation [S2]:

$$k^n \chi(k) = S_0^2 \sum_i N_i F_i(k_i) k_i^{n-1} e^{-2\sigma_i^2 k_i^2} e^{-\frac{2r_i}{\lambda_i(k_i)}} \times \frac{\sin[2k_i r_i + \varphi_i(k_i)]}{r_i^2} \quad (1)$$

where $S_0^2$ is the EXAFS reduction factor due to many-body effects. $N_i$ is the number of neighbouring atoms at a distance $r_i$, $F_i(k_i)$ is the backscattering amplitude, $k_i$ is the wave number of the photoelectron, $\sigma_i^2$ is the EXAFS Debye-Waller factor measuring the mean square relative displacements (MSRDs) of the photoabsorber-backscatterer pairs due to their thermal motion and $\lambda_i$ is the photoelectron mean free path. The $\varphi_i$ is the phase shift implemented in the backscattered wave. Here, fitting of the EXAFS spectra in the context of single scattering approximation (eq. 1) was performed by employing the crystal model of the layered $\beta$-FeSe [S3]. The local structure is fitted within the first coordination shell around the Se absorber which involves 4 Fe atoms at a distance ~2.39 Å. Fitting the data with the single shell requires refining four parameters: the amplitude reduction factor ($S_0^2$), the threshold energy ($E_0$), the Fe-Se distance ($r_i$) and their corresponding thermal parameter ($\sigma_i^2$). However, the availability of temperature-dependent EXAFS data allows to keep $S_0^2$ and $E_0$ constant by evaluating them over the broad temperature range and then taking their average. Therefore, fitting of the temperature-dependent Se K-edge EXAFS was performed by refining only the Fe-Se $r_i$ and $\sigma_i^2$ with maximum number of independent fitted parameters to be ($2\Delta k\Delta R/\pi$) around 7.

The MSRDs for a pair of atoms reflect the lattice configurational disorder expressed through the sum of temperature-independent ($\sigma_s^2$; static) and temperature-dependent ($\sigma_d^2$; dynamic) terms, $\sigma^2 = \sigma_s^2 + \sigma_d^2(T)$ [S4]. Phenomenological models like the correlated Einstein or the correlated Debye model can be fitted against the MSRDs [S4,S5]. The correlated Einstein model approximates the projected vibrational density of states with a single resonant frequency—a δ function centered around the effective vibrational frequency of a given path:

$$\rho_R(\omega) = \delta\big(\omega - \omega_E(R)\big) \quad (2)$$

hence, the temperature-dependent term takes the simple form:

$$\sigma_d^2(T) = \frac{\hbar^2}{2\mu k_B \theta_E} \coth\left(\frac{\theta_E}{2T}\right) \tag{3}$$

where $\mu$ is the reduced mass of the respective atom pair and $\theta_E$ is the Einstein temperature, related to the Einstein frequency ($\omega_E = k_B \theta_E / \hbar$). The correlated Debye model gives a spherical approximation to $\sigma^2$ with the projected vibrational density of states for an atomic bond:

$$\rho_R(\omega) = \frac{3\omega^2}{\omega_D^3}\left(1 - \frac{\sin(\omega R/c)}{\omega R/c}\right) \tag{4}$$

where $\omega_D = k_B \theta_D / \hbar$ is the Debye frequency and $c = \omega_D / k_D$ where $k_D = (6\pi^2 N/V)^{1/3}$ and $N/V$ is the atomic density number of the crystal.

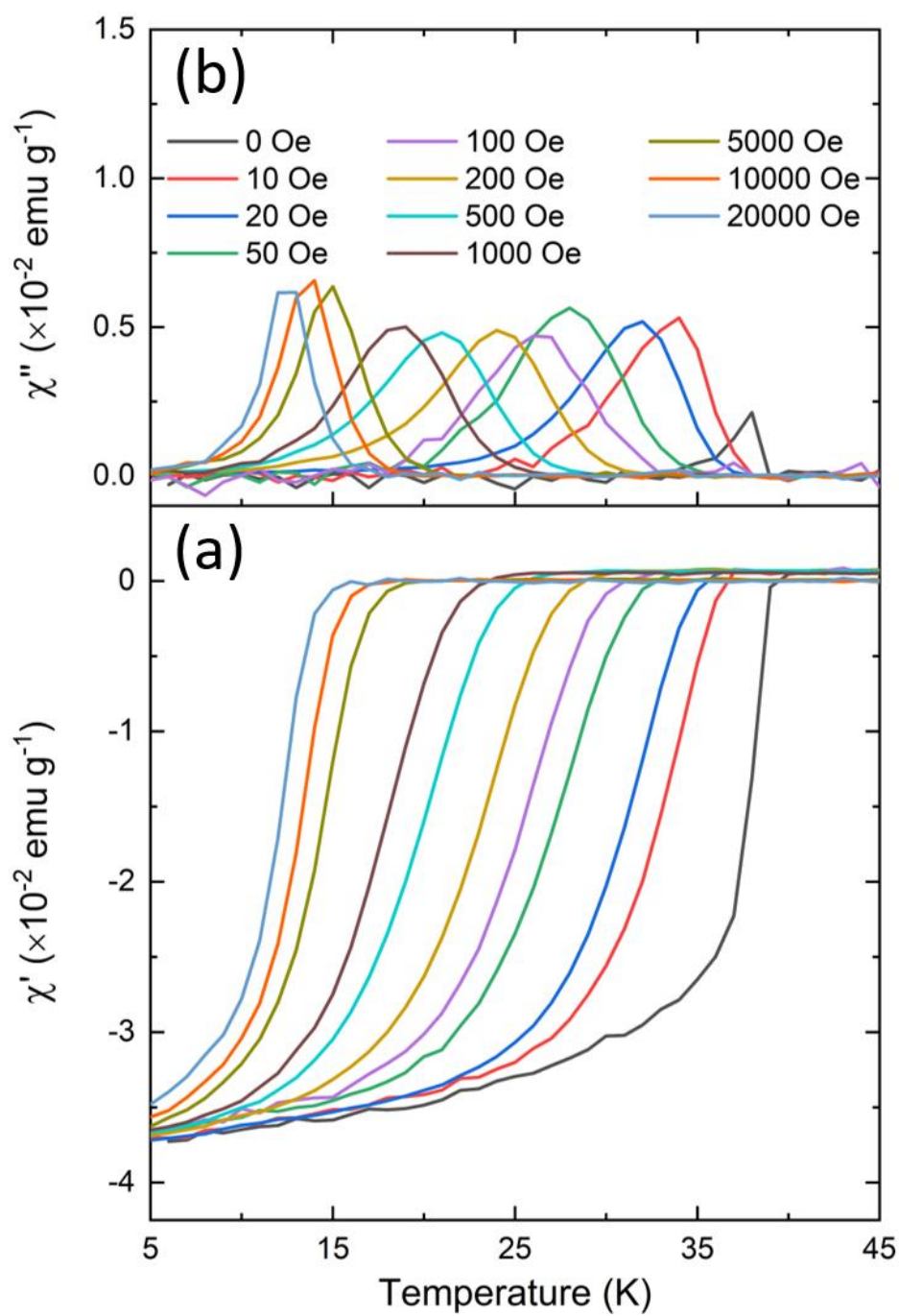

**Figure S1.** (a) real ($\chi'$) and the (b) imaginary ($\chi''$) parts of the AC susceptibility ($H_{ac}$ = 1 Oe and $f$ = 499 Hz) for the deuterated, annealed Li$_{1.0}$(C$_5$D$_5$N)$_y$Fe$_{2-z}$Se$_2$ compound under various DC magnetic fields.

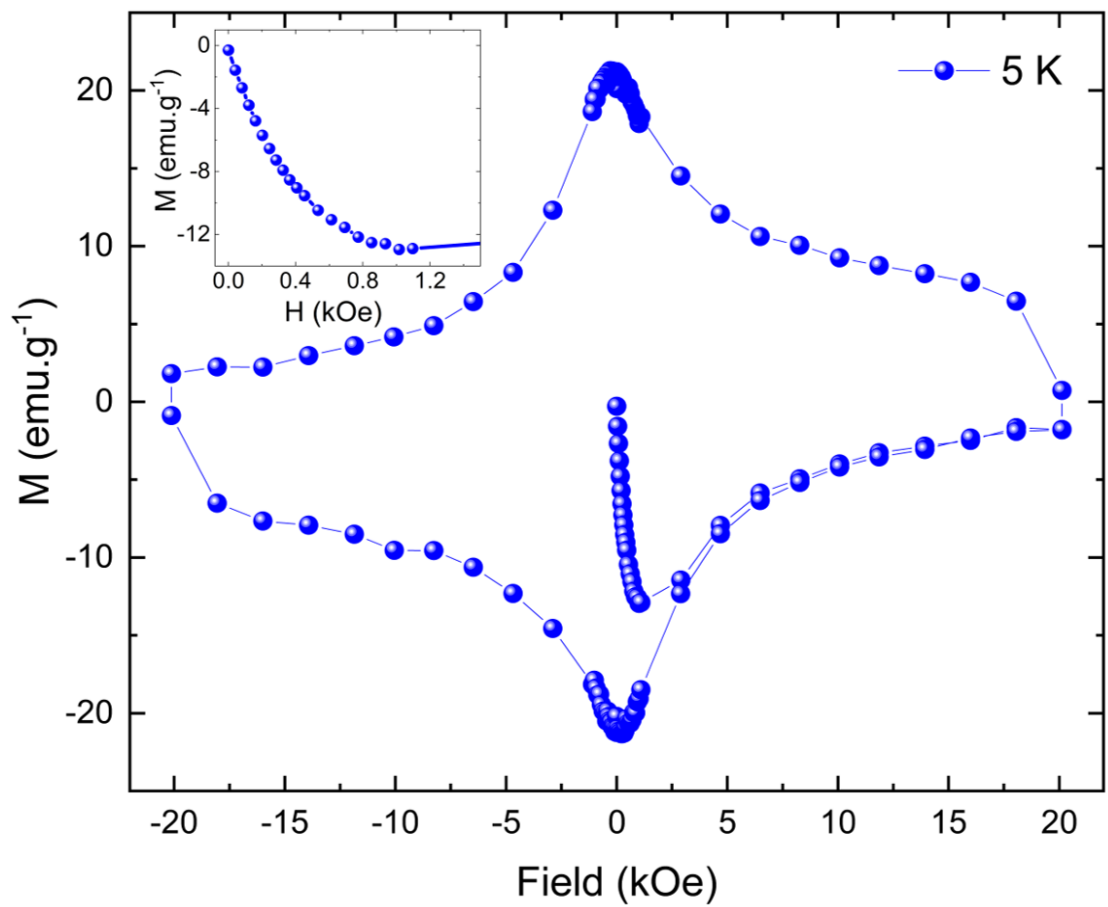

**Figure S2.** The magnetisation against DC magnetic field (*M-H*) loop of the deuterated, annealed Li$_{1.0}$(C$_5$D$_5$N)$_y$Fe$_{2-z}$Se$_2$ material, at 5 K. The inset shows the low-field region.

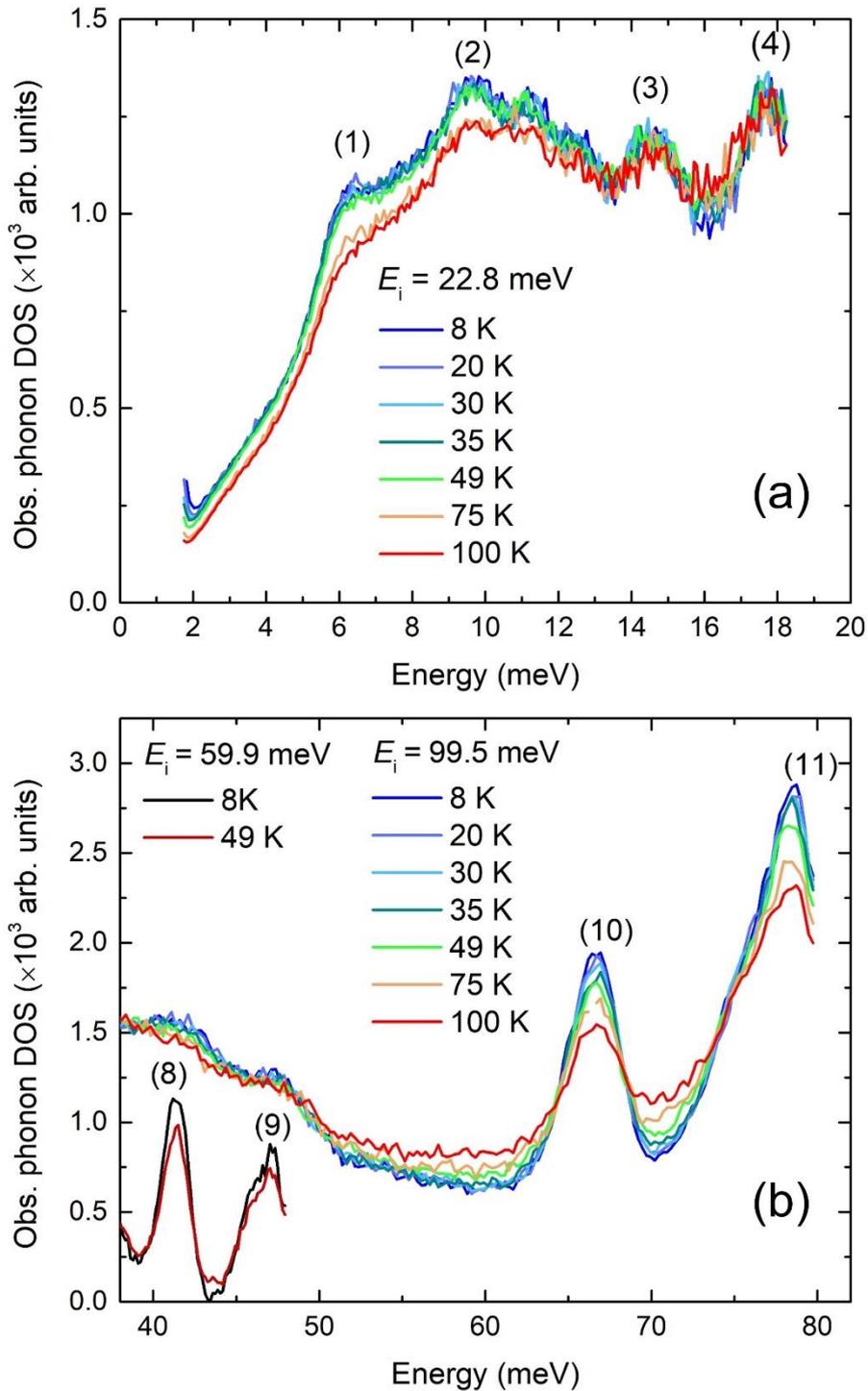

**Figure S3.** The measured PDOS of the annealed $Li_{1.0}(C_5D_5N)_yFe_{2-z}Se_2$ as a function of temperature, derived from spectra recorded with incident energy: (a) $E_i = 22.8$ meV. Features #1 and #2, better resolved at low temperatures, are attributed to TA phonon modes of equal mixture of Fe and Se vibrations (*cf.*, main text). They become broaden above the $T_c$ ($\cong 39$ K) and at elevated temperatures (75 – 100 K). (b) $E_i = 59.9$ meV, reveals features #8, #9 above and below $T_c$, and with $E_i = 99.5$ meV, shows the evolution of the modes #10, #11 in the temperature range 8 – 100 K. These features assign well to the vibrational states of the intercalated pyridine ($C_5D_5N$).